\documentclass[12pt,preprint]{aastex}

\newcommand{\beq}{\begin{equation}}
\newcommand{\eeq}{\end{equation}}

\newcommand{\gsim}{\gtrsim}

\begin{document}

\title{Search for Galaxy Clustering around a Quasar Pair at z=4.25 found in
the Sloan Digital Sky Survey}

\author{Masataka Fukugita\altaffilmark{2},
Osamu Nakamura\altaffilmark{2},
Donald P. Schneider\altaffilmark{3}
Mamoru Doi\altaffilmark{4},
and
Nobunari Kashikawa\altaffilmark{5}
}
\altaffiltext{1}{Based on data collected at Subaru Telescope, which is operated by the National Astronomical Observatory of Japan.}
\altaffiltext{2}{Institute for Cosmic Ray Research, University of Tokyo,
Kashiwa 277 8582, Japan; fukugita@icrr.u-tokyo.ac.jp, 
nakamura@sdss1.icrr.u-tokyo.ac.jp}

\altaffiltext{3}{Department of Astronomy and Astrophysics, Pennsylvania 
State University, University Park, PA 16802, U. S. A.; 
dps@astro.psu.edu}

\altaffiltext{4}{Institute for Astronomy, University of Tokyo, 
Mitaka, 181 8588 Japan}

\altaffiltext{5}{National Astronomical Observatory,
Mitaka, Tokyo 181 8588, Japan}

\begin{abstract}

We present an investigation of the environment ($\approx$ 600 kpc radius)
of a pair of luminous $z=4.25$ quasars, SDSS J1439$-$0034 A,B,
separated by 33$''$.
An analysis of high-quality Subaru spectra of the quasars suggests that
this configuration is indeed a physical pair and not a gravitational lens;
the redshifts are slightly different (although marginally consistent with
being the same), and the two spectra have strikingly different features.
We search for bright galaxies ($L\gsim 0.4 L^*$) having similar redshifts 
using the V dropout technique 
and semi-narrow band imaging looking for strong Ly $\alpha$ emission.
We find no enhancement in the galaxy density around the quasar pair;
its environment differs very little from a general field, with the upper
limit of the density enhancement being about 3 at a 90\% confidence. We
infer that bright quasars happened to appear in two normal galaxies in
a general field.  
\end{abstract}

\section{Introduction}

Luminous quasars are rare phenomena in the Universe. When the density 
of luminous quasars reaches a maximum at 
$z\approx 2$, its comoving density is about $10^{-7}$ times the density of 
galaxies (e.g., Schmidt, Schneider \& Gunn 1995). 
High-redshift quasars are often used as signposts to 
search for accompanying galaxies, 
as one expects a rich environment around quasars 
(e.g., Hall \& Green 1998; Djorgovski 1998; 1999 and references therein). 
Pairs of quasars may be taken as particularly promising markers 
of high-density regions, since the probability that two quasars 
exist in a small volume is very small, unless they are embedded in a special
environment such as rich clusters or protoclusters of galaxies. 
Therefore, the search for physical pairs of quasars, especially at high
redshifts, is particularly interesting from the
point of view that they may permit exploration of an 
unusually rich environment 
in the early universe. 
Taking advantage of wide-field multicolour CCD imaging
of the Sloan Digital Sky Survey (SDSS; York et al. 2000), 
Schneider et al. (2000) discovered
a pair of  luminous quasars at $z=4.25$ separated by $33''$
in their follow-up spectroscopy at the Hobby-Eberly Telescope.
In this paper the two quasars will
be designated as SDSS J1439$-$0034 A, B, with A being the brighter component.
Spectroscopic features indicated that this pair is most likely to be a
physical association, not a gravitational lens, although
the spectra were of limited S/N.
The projected
distance is $0.16h^{-1}$ Mpc with a non-zero lambda ($\Omega_0=0.3$,
$\lambda=0.7$) cosmology. This system's separation is among the
smallest known at high redshifts. No comparable high-redshift
quasar pairs have
been found among the 16,713 quasars 
in the 1360 deg$^2$ given 
in the SDSS First Data Release (Schneider et al. 2003). 

The widely-accepted model of galaxy formation based on the
cold dark matter (CDM) dominance suggests that luminous
quasars may be embedded in rich
environment (Martini \& Weinberg 2001; Haiman \& Hui 2001).
In CDM models, quasars are usually 
ascribed to a phenomenon associated with 
very rare, high peaks of Gaussian fluctuations (Efstathiou
\& Rees 1988). Such fluctuations preferentially reside in 
the region where one expects rich clusters as the Universe evolve.
This is particularly true if the lifetime of individual quasars
is long, and only very rare peaks are associated with luminous
quasars. 

With the anticipation that the environment of two 
luminous quasars would be particularly rich, we investigate  
the distribution of galaxies around the SDSS quasar pair to determine
whether this system is embedded in a rich protocluster of galaxies.
We have carried out deep $V$, $R$ and $I$ imaging and semi-narrow band
imaging with the $N$642 filter, which was available at the observatory, 
for the $6'$ field around 
the SDSS quasar pair using the Subaru Telescope at Mauna Kea. 
Our prime purpose is to study the spatial distribution of
galaxies that show Lyman break features (Steidel, Pettini \& Hamilton 1995)
in the $V$ passband (Madau et al. 1996), which indicate $z\approx 4$, and
those that show Lyman $\alpha$ emission around the same
redshift.  
Additional high-quality spectroscopic observation was made to study 
whether the two quasars are lensed images or are two distinct
objects that have the same redshift.

\section{Redshifts of the pair of quasars}

An important issue to decide is whether the quasars are two gravitationally
lensed images
of the same source or two physical distinct objects.  Schneider et al. (2000)
concluded that the large separation (33$''$) 
and the apparent differences in the
spectra strongly suggested that this is a pair of quasars and not a
gravitational lens.  
Given the limited S/N of the spectra, however, 
this conclusion can only be regarded as tentative.

To improve upon the spectra obtained by Schneider et al. (2000)
spectroscopic observations of the quasar pair were made on 20 June 2001 using
FOCAS at the Subaru Telescope (Kashikawa et al. 2002).
The instrument configuration was a $0.''4$ long-slit,
a  300 line mm$^{-1}$ grism blazed at 5500\AA~(B300 grism),
and an SY 47 blocking filter. The 3600 sec
exposure was made with the slit
rotated to include the two quasars. 
FOCAS CCD detectors produce
an image scale of $0.''104$ pixel$^{-1}$ and a dispersion of 
1.40 \AA~pixel$^{-1}$. The pixels are binned 3$\times$2 
(spatial$\times$spectral) during readout. 
The spectrum covered 4700 \AA~ to 9400 \AA~
at a resolution of approximately $\lambda/\Delta \lambda\simeq 1000$. 
Observing conditions were reasonable, although some cirrus was recognized;
the seeing was $0.95''$ FWHM. The spectrophotometric
calibration was made with BD+28$^\circ$4211. 
Absolute spectrophotometry was obtained by adjusting the scale to 
the broad band $R$ magnitude for each quasar.

The Subaru spectra of SDSS J1439-0034 A,B are displayed in Figure 1.  Both
spectra show the standard features of high-redshift quasars: a significant
Ly $\alpha$ forest, a strong, highly absorbed Ly $\alpha$ emission line,
prominent C IV emission line, and other weaker features.  The redshifts of
the quasars, as measured from a weighted value of the emission lines other
than Lyman $\alpha$, are 4.228 $\pm$ 0.005 (A) and 4.243 $\pm$ 0.009 (B).
The redshifts are marginally consistent with a single value.  The velocity
difference is 860 km s$^{-1}$, with an uncertainty of comparable size.

The spectra show a number of striking differences, in particular in the
strength of the C IV line (the equivalent width in B is nearly twice that
in A), and the presence of a strong N V/O I/Si II/C IV associated absorption
system in the spectrum of B.  The redshift of the absorption system,
4.249 $\pm$ 0.002, is slightly larger (340 km s$^{-1}$) than the emission
line redshift of B; this is consistent with the appearance of the C IV line
profile.

We conclude that all of the available data support the interpretation of
SDSS J1439-0034 A,B as a physical pair separated by several hundred kpc.

\section{Imaging Observation and Data Reduction}

The imaging observation was made on 19 and 20 June, 2001 also using 
FOCAS at the Subaru Telescope. The circular field of view has a 
radius of $3.0'$, and the image scale is $0.104''$ pixel$^{-1}$
on two 2k$\times$4k CCD. 
A small fraction of the field is obscured by the autoguider. 
The conditions were photometric, when the VRI photometry 
was acquired on the first night,
but light cirrus was present on the second night when the narrow band imaging
was obtained.  The image quality was approximately $0.7''$
FWHM for both nights.  The total exposure times 
were 6000s, 3600s, 2340s for $V$(5500), $R$(6600) and $I$(8050), 
and 3600s for $N$642,
whose pass band extends from 6364 to 6492 \AA.
Data were processed using a standard IRAF photometry package, and
objects were identified with SExtractor version 2.2.1. 
The objects are catalogued if they meet a 2$\sigma$
threshold after Gaussian smoothing with the width of 
$\sigma=3$ pixels 
in the $R$ band frame.
This detection threshold corresponds to $R_{\rm AB}\approx 26$,
or roughly $I_{\rm AB}\approx 25.5$. 
(All magnitudes reported in this paper are on the
AB magnitude scale, and throughout the
rest of this paper we will drop the AB subscript.) 
The characteristic magnitude 
of $z\approx 4$ Lyman break
galaxies is $I^*\simeq 25$ derived by Steidel et al. (1999).
Our depth with FOCAS is shallower by about 1 mag
than that of Ouchi et al. (2001), who studied 
$z\approx 4$ galaxies using the prime focus imager
of the Subaru Telescope; 
our observations, however, are sufficiently deep 
to detect significant numbers of
bright galaxies at $z\approx 4$, as our simulation (described later)
indicates.
The photometric zero points determined from standard stars agree with 
those of SDSS photometry to within 0.1 mag after the colour transformation.
The FOCAS field that contains quasars will be referred to as Field Q. 
The basic properties of the quasars are listed in Table 1.

Extra $VRI$ imaging observations were made with slightly shorter exposure
times for a `blank' field centred at
$\alpha_{2000}=23^h$24$^m$36$^s$ and $\delta_{2000}=0^\circ$04$'$30$''$; 
Field B); this will be designated as Field B.
 Note that the FOCAS field is sufficiently wide that the majority
of Field Q can also be used as a background field.
The blank field increases statistics of background galaxies.

\section{Distribution of High Redshift Galaxies around the Quasar Pair}

The catalogue contains 1181 objects fainter than $I>23.0$ in Field Q
(effective area after excising the masked area for bright stars and
a satellite track is 21.6 sq. arcmin), and 954 objects in Field B
(21.6 sq. arcmin).
We apply the $V$ dropout technique to
select candidates of $z\sim 4$ galaxies.
We use the population synthesis model
of Kodama \& Arimoto (1997) to calculate the evolution
tracks in colour space, but we have checked that the tracks of 
GISSEL95 (see, Bruzual \& Charlot 1993) differ little for the redshift of
interest here. The effects of 
Ly $\alpha$ absorbers are incorporated according to
Madau (1995) (see also Steidel et al. 1999).
A variety of star formation histories is adopted to calculate the
location for $z\approx 4$ galaxies with internal reddening varying from 
$E(B-V)=0$ to 0.3, and the region in 
$V-R$ vs $R-I$ space is optimised using a Monte Carlo simulation which
we describe in what follows. The condition we adopt to select
galaxies centred at $z\approx 4.24$ galaxies is 
\begin{eqnarray}
0.95<V-R<2.0,\cr
~~~~~~~~~~0.59(R-I)+0.54< V-R< 3.6(R-I)+0.4,\cr
23.5<I,~~~~~~~R-I<1.0.
\label{eq:colcond}
\end{eqnarray}
This colour-colour region (delineated with the dashed line) and 
typical evolution tracks are shown in Figure 2, together
with the galaxies we detected on one of the 2k$\times$4k CCD exposed for
Field Q (we omitted data points which have large photometric errors,
$\delta I>0.07$ mag in the I band: largely scattered data points in the
$R-I<0$ region are mostly ascribed to poor photometry in the $I$ band). 
The tracks cover the range from 
a short, single burst ($\tau=0.1$ Gyr) to continuous star
formation to explore wide star formation histories and a range of
$E(B-V)$ specified above. The formation redshift
is assumed to be $z=10$, but tracks with lower formation redshift pass
through the same region in colour space.

With condition (\ref{eq:colcond}) 
we obtain 26 galaxies in Field Q and 18 in Field B.
Those galaxies that satisfy (\ref{eq:colcond}) in Field Q 
are plotted by large circles, together
with all others with $23.5<I$ which are denoted by dots, in Figure 3.  

A Monte Carlo simulation is carried out using the colour, magnitude and 
redshift distribution of galaxies in HDF-N 
(Fern\'andez-Soto, Lanzetta \& Yahil 1999) to find the selection criterion,
and to estimate the contamination and completeness of the sample, 
as was done in Ouchi et al. (2001).
Figure 4 presents an example of Monte Carlo simulation, 
showing the redshift distribution of the galaxies selected with 
(\ref{eq:colcond}). Taking our window to be in the range $4.0<z<4.6$,
the contamination from low redshift objects and those just below or
above the window is 60\% for $I=24-25.5$. Almost all 
objects with $I<23.5$ are of low redshifts. 
The completeness is about 75 \% for $I=23.5-25$, but drops to
60\% for $I\approx 25$.

It is already clear in Figure 3 that there is no recognizable clustering 
of galaxies around the quasar pair. To make this statement more precise,
we show in Figure 5 number counts of galaxies in the $I$ band for those
in a circular region (indicated in Figure 4) of the radius $90''$ 
(physical distance: 0.61 Mpc for $h=0.7$) from the centre of the quasar pair, 
and those in other parts of 
Field Q plus Field B. The choice of
a $90''$ radius, which was rather arbitrary, was made to 
maximize the number of galaxies encircled 
to derive a conservative limit on the surface density of 
galaxies around the quasars.
The area for the quasar field is 5.8 sq. arcmin,
and that of the background totals 37.3 sq. arcmin.
We applied the above-mentioned corrections for incompleteness
and contamination. The dotted triangles are number counts for
$z\approx4$ galaxies obtained by
Ouchi et al. (which agree with those of Steidel et al. 1999)
from their survey of a much wider field. The number counts in
both regions agree for all magnitudes brighter than $I=26$,
and with Ouchi et al. counts for $I<25$. The latter disagreement beyond
$I>25$ indicates incompleteness of our catalogue, but this completeness
is not important for our purpose.
This figure confirms that there is no enhancement of galaxies 
with the luminosity $L>0.4L^*$ in the
vicinity of the quasar pair.
From Poisson statistics we derive an upper limit of the density
enhancement around the quasars to be $\rho_{\rm qso}/\bar\rho<3$ at
the 90\% confidence level for the 90$''$ circular region.
Note that this calculation includes two quasars as bright galaxies.

Our final discussion concerns $N$642 semi-narrow band imaging. The
passband of the filter covers Ly $\alpha$ emission for $4.23<z<4.33$,
but the band width is not narrow enough for sensitive search for Ly
emission. So, we adopt a rather weak criterion that $N-I<1.0$
to select `$N$ bright' galaxies. Among the high $z$ galaxy candidates 
that meet (\ref{eq:colcond}) those that
satisfy this criterion are shown by solid circles in Figure 3 above. The
distribution of these `$N$ bright' galaxies are scattered, no
clustering is seen around quasars. Incidentally, we have confirmed
by removing the $I>23$ mag condition 
that there is no quasars in Field Q which fall in this redshift range other
than the two that are already known.

\section{Conclusions}

We have shown that the field in the vicinity of the quasar pair
SDSS J1439$-$0034 at $z=4.24$ hardly differs from the general 
filed regarding the surface density of bright galaxies at around
$z\approx 4$. A density enhancement, if present, is no more than
three times the field density at a 90\% confidence. 
We consider that this is rather unexpected.

Although this is but one example (though the pair of quasars make it
a particularly attractive one), our null results suggest that luminous
quasars are not always signposts for high-density regions. This 
is perhaps an example of a case where two normal 
galaxies happened by chance to simultaneously
display the quasar phenomenon.
The scarcity of luminous, high redshift quasars is not solely
(or even primarily) due to the requirement that they reside in the
rare, high density perturbations; a relatively short lifetime of the
quasar phenomenon at high redshift could play an important role.

\vskip5mm\noindent

We would like to thank Youichi Ohyama for his very efficient help for
our observation at the Subaru Telescope. MF is also grateful to Richard
Ellis and the Astronomy Group of Caltech for their hospitality while this
article was being completed. MF is supported in part by Grant in Aid
of the Ministry of Education (No 15204011), and DPS by the NSF Grant 
(AST03-007582).

\clearpage

\begin{figure}
\plotone{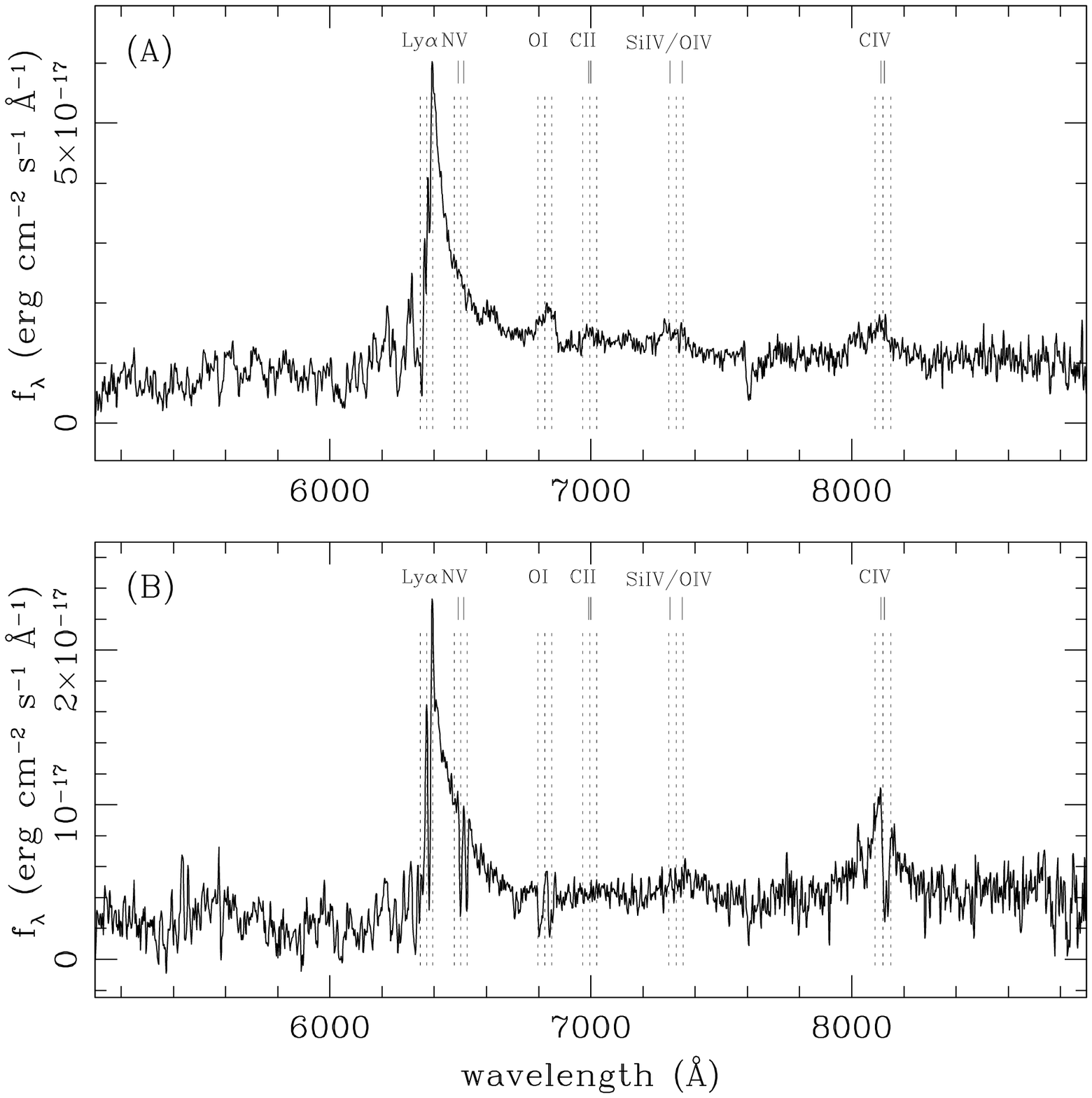}
\caption{
Subaru spectra of a pair of quasars SDSS J1439$-$0034 A, B. The spectral
resolution is approximately 1000.
The positions 
of typical emission lines are drawn for assumed redshifts of $z=4.22,
4.24$ and 4.26. Short slabs at the top of dotted lines indicate the
position of doublets NV, CII, SiV and CIV for  $z=4.24$. } 
\end{figure}%

\begin{figure}
\plotone{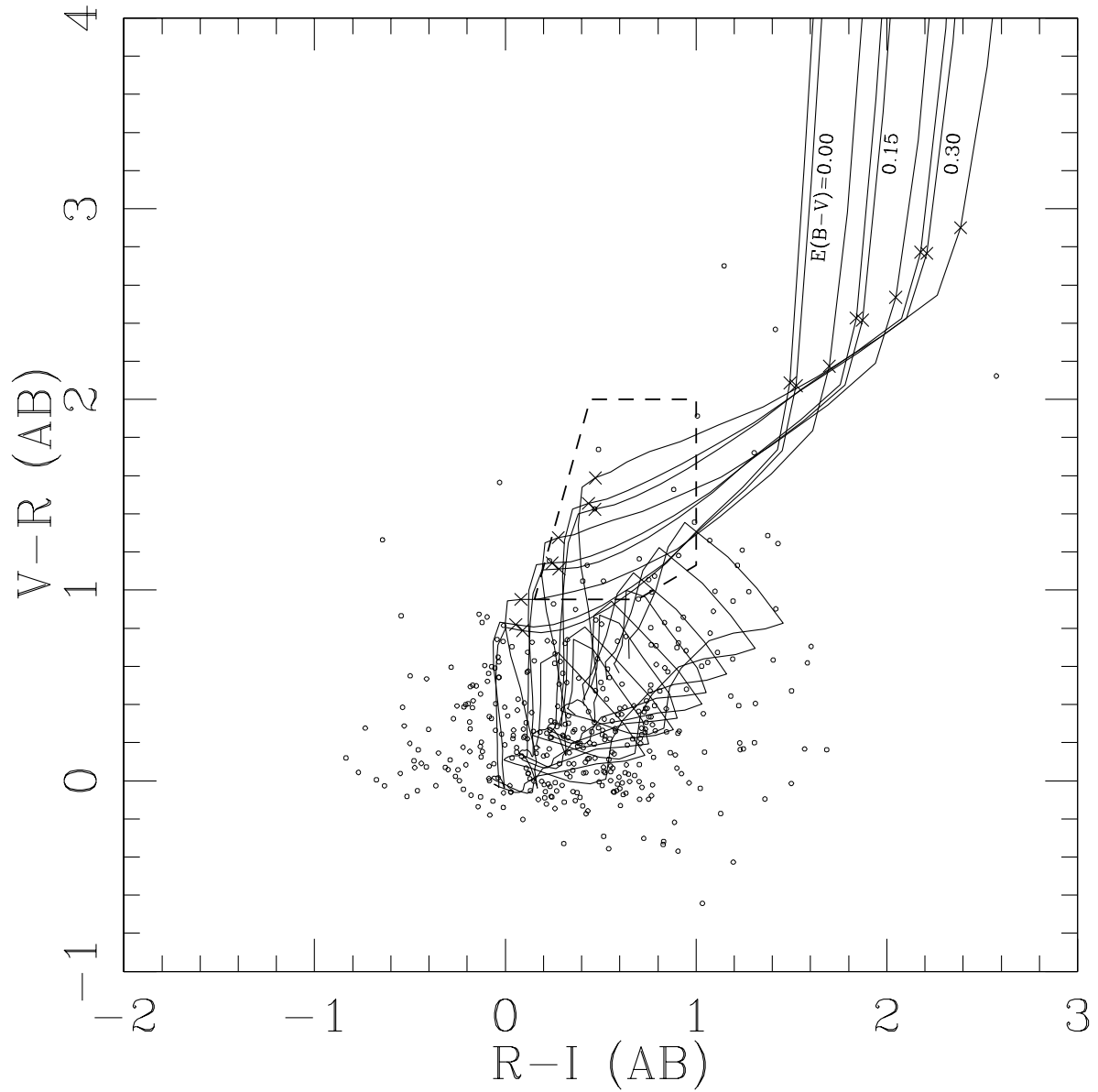}
\caption{Galaxies in the $(V-R)_{\rm AB}$ versus $(R-I)_{\rm AB}$ 
plane (the galaxies are taken
from a subsample of Field Q). 
The evolution tracks
shown correspond models with a short burst at a high redshift 
to continuous star
formation with extinction $E(B-V)=0-0.3$. The crosses on the tracks show
the position at $z=4$ and 5.  The area delineated by the dashed line is
condition (1) to select $z\approx 4.24$ galaxies.
}

\end{figure}%

\begin{figure}
\plotone{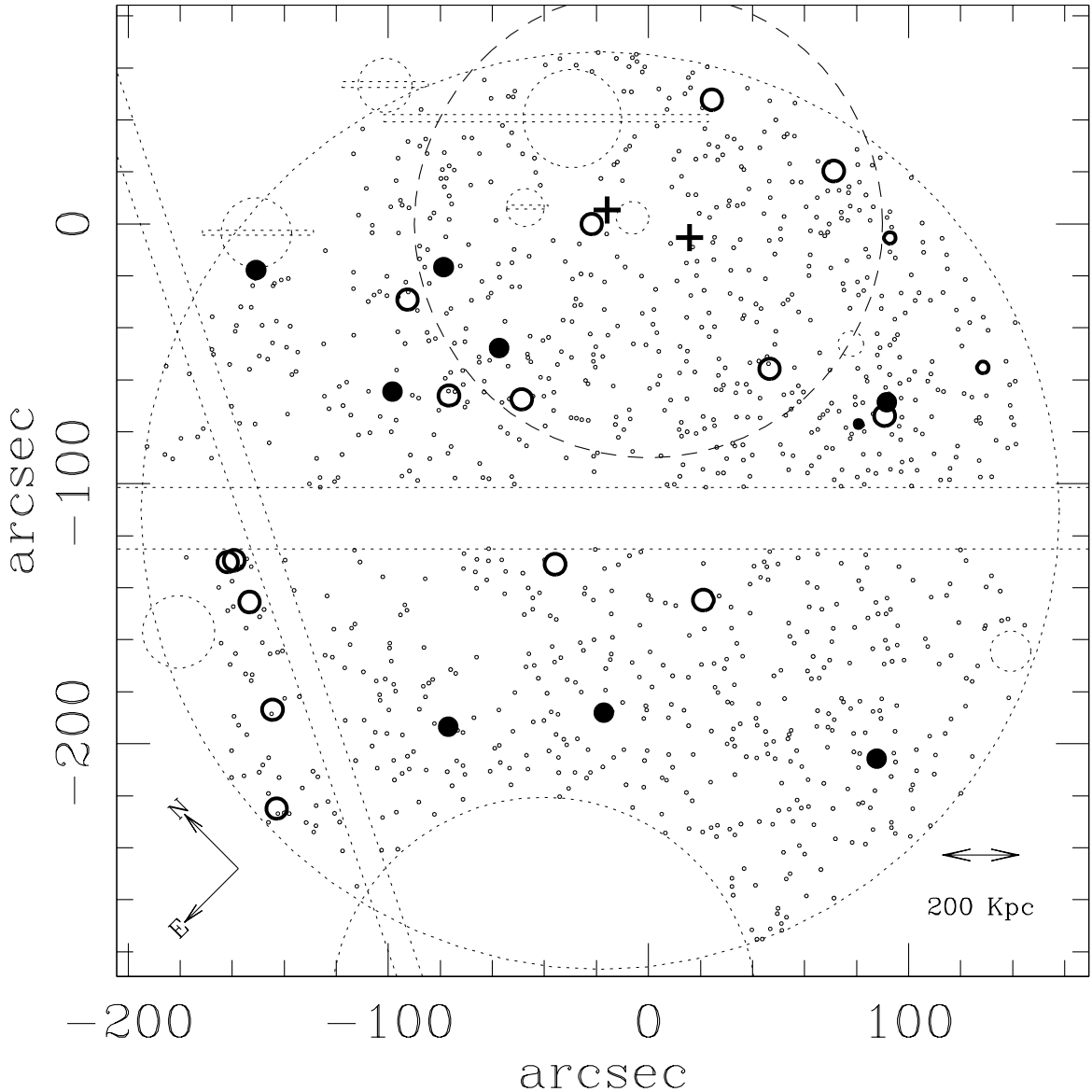}
\caption{
Distribution of galaxies detected with the 2$\sigma$ threshold in 
Field Q. Large circles show the galaxies that satisfy condition 
(1), yet brighter than $I_{\rm AB}=26$  (those fainter than 
$I_{\rm AB}=26$ are shown with smaller circles).
Among them $N642$-passband bright galaxies are shown with solid
circles. The two crosses denote the pair of quasars SDSS J1439$-$0034 A, B
(A is at the left). Dotted boundaries show the masked region (bright
stars or satellite tracks) and a gap between the two CCDs. The dashed
circle is 90$''$ (0.61 Mpc for $h=0.7$) from the centre of 
the quasar pair, which is also
taken to be the origin of the coordinates.} 
\vspace{0.5cm}
\end{figure}%

\begin{figure}
\plotone{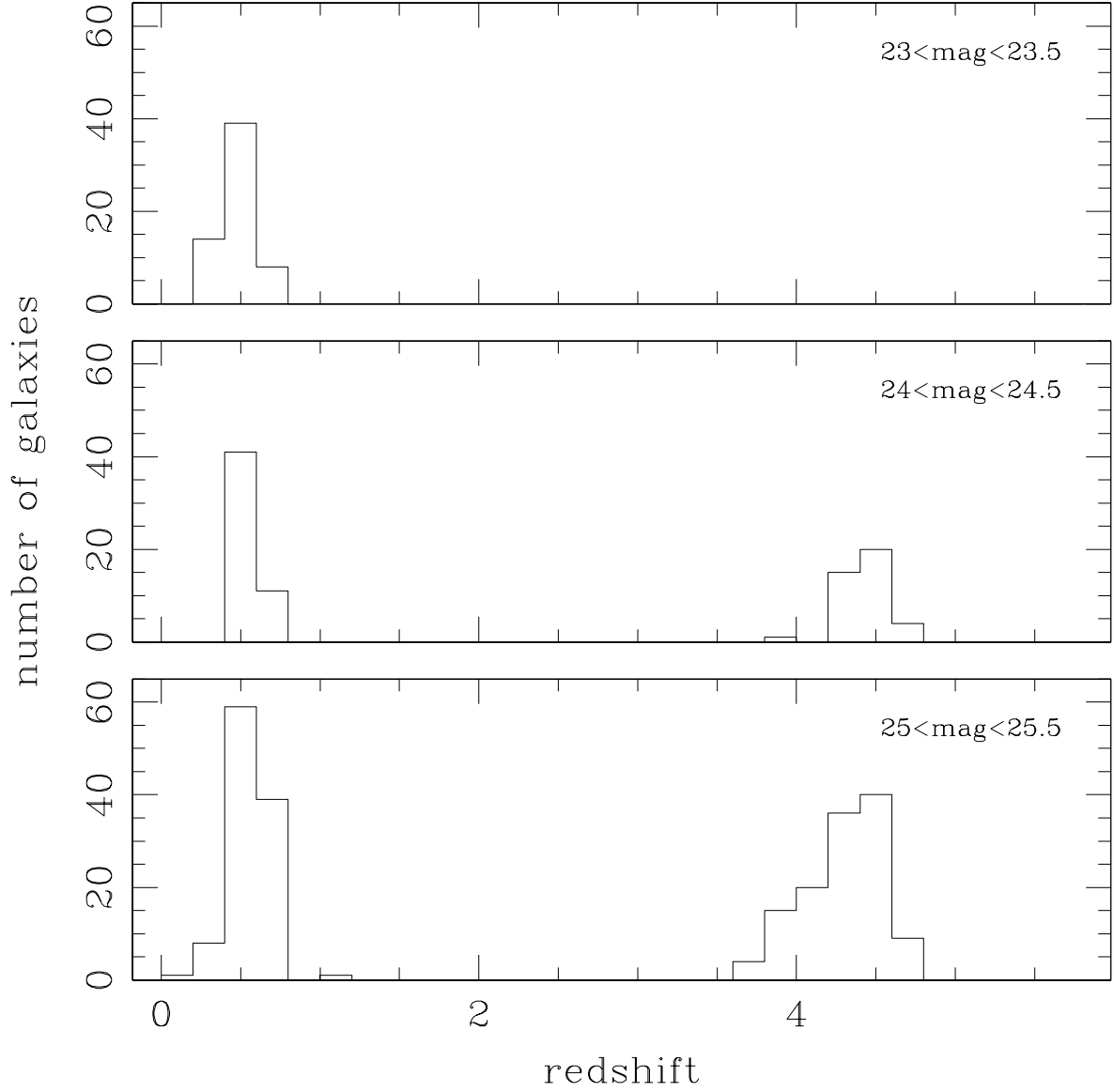}
\caption{
Redshift distribution of galaxies that satisfy condition (1) expected
from a Monte Carlo simulation using the empirical galaxy data from
HDF-N (Fernandez-Soto et al. 1999).}
\end{figure}%

\begin{figure}
\plotone{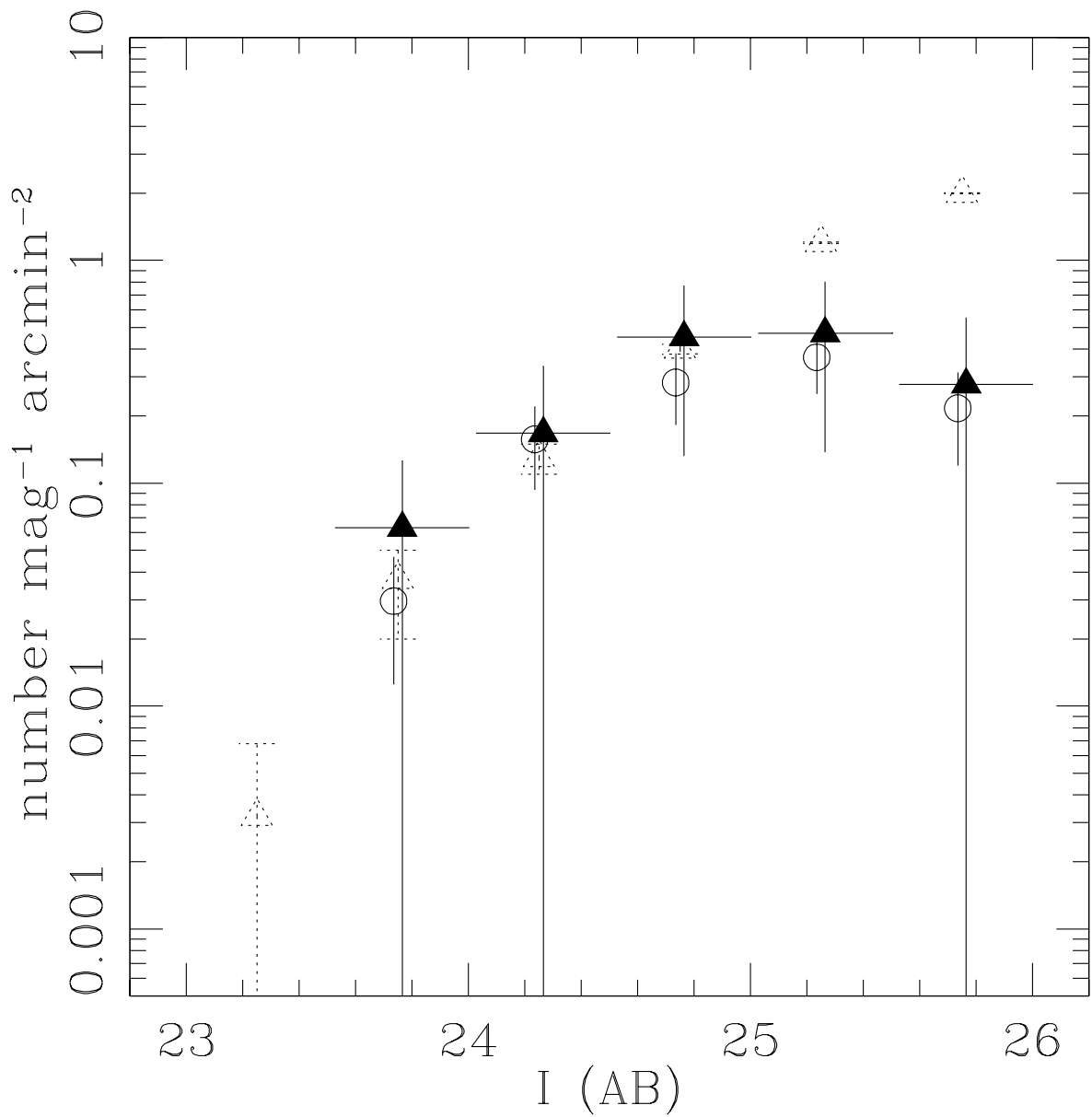}
\caption{
Number counts of galaxies in the $I_{\rm AB}$ band 
expected for $4.0<z<4.6$ after the corrections
for completeness and contamination as described in the text.
Those in the QSO field  (90$''$ from the centre of the quasar pair)
are represented by solid triangles, and those in the background fieled
(Field Q excluding the QSO field and Field B combined) are indicated by
open circles. The dotted triangles show number counts of Ouchi et al. 
(2001).} 
\end{figure}%

\clearpage

\begin{table*}
\begin{center}
\caption{Pair Quasars SDSS J1439$-$0034}
\begin{tabular}{lccccccc}
\tableline\tableline
  &  RA(J2000)  & DEC(J2000)  & $V_{\rm AB}$  &  $R_{\rm AB}$  & $I_{\rm AB}$ & $N642_{\rm AB}$ & redshift \\
\tableline
A & 14$^h$39$^m$52.58$^s$ & $-$00$^\circ$ 33$'$59.2$''$ & 
       21.65  & 20.47  &  ---$^{(a)}$  &  19.88  &  4.228$\pm$0.005  \\
B & 14$^h$39$^m$51.60$^s$ & $-$00$^\circ$ 34$'$29.2$''$ & 
       22.76  & 21.77  &  21.61  &  20.95 & 4.243$\pm$0.009  \\
\tableline
\end{tabular}
\end{center}
(a) The $I$ band signal of quasar A is saturated in our image.
\end{table*}

\end{document}